\documentclass[letterpaper]{article} 
\usepackage[preprint]{aaai2027} 
\usepackage[hyphens]{url} 
\usepackage{graphicx} 
\usepackage{natbib} 
\usepackage{caption} 
\usepackage{amsmath,amssymb}
\usepackage{booktabs}
\usepackage{multirow}
\usepackage{xcolor}
\usepackage{tikz}
\usetikzlibrary{arrows.meta,positioning}
\title{Socrates-RAG: Premise-Directed Inquiry\\
against Coordinated Evidence Poisoning}

\author{
Renyu Zhao\corresponding\textsuperscript{\rm 1},
Xinyuan Zou\textsuperscript{\rm 2},
Lanbin Liu\textsuperscript{\rm 2}
}
\affiliations{
\textsuperscript{\rm 1}Tencent SSV, Beijing 100193\\
\textsuperscript{\rm 2}School of Future Cities, University of Science and
Technology Beijing, Beijing 100083\\
roizhao@gmail.com, M202410089@xs.ustb.edu.cn, liulanbin@ustb.edu.cn
}

\begin{document}
\maketitle

\begin{abstract}
Retrieval-augmented generation (RAG) defenses typically decide how to filter
or aggregate a fixed retrieved set. In open-corpus question answering,
however, decisive evidence may be absent from the initial context but
retrievable, making the \emph{next query} part of the reliability problem. We
introduce Socrates-RAG, a premise-directed active retrieval policy that
represents competing answers, selects an unresolved premise whose resolution
would discriminate them, and uses newly acquired evidence to refine a
subsequent query before answering or abstaining. We formalize the resulting
finite-budget evidence state and give a conditional rescue guarantee relative
to repeated or topical-query policies.

We evaluate Socrates-RAG against a matched control in which the same backbone
generates ordinary relevance-oriented search queries; both policies share the
initial evidence, deterministic retriever, two-query/top-three budget, answer
prompt, and label-free evidence-chain release rule. On a disjoint 48-world
counterfactual evaluation, premise-directed inquiry raises safe accuracy from
79.2\% to 93.8\%, with 8 wins, 1 loss, and 39 ties (two-sided exact
$p=.0391$). Decisive-evidence recall improves by the same margin, while unsafe
answers fall from one to zero. Both policies solve all 24 one-hop cases; the
gain is concentrated in two-hop cases, where Socrates-RAG substitutes a newly
resolved premise into its second query. This controlled study isolates a
specific benefit of premise-directed acquisition without claiming general
robustness on the open Web.
\end{abstract}

\section{Introduction}
\label{sec:intro}

RAG grounds generation in retrieved evidence rather than relying only on
parametric memory~\cite{lewis2020rag,karpukhin2020dpr,gao2023ragsurvey}.
The same interface creates a security boundary: an attacker who can add
or manipulate retrieved content can influence both the evidence shown to
the model and the answer generated from it.  Corpus poisoning has been
shown to steer RAG toward attacker-chosen answers with only a small number
of optimized passages~\cite{zou2025poisonedrag}; malicious retrievers can
also expose downstream generators to injected instructions
\cite{clop2024backdoored}.

Many defenses operate on a fixed set.  TrustRAG filters embedding-space
anomalies~\cite{zhou2025trustrag}; RobustRAG isolates passages and aggregates
their answers with certified guarantees under its threat model
\cite{xiang2024robustrag}; ReliabilityRAG uses retrieval rank and answer
contradictions~\cite{shen2025reliabilityrag}.  These mechanisms answer
\emph{which retrieved evidence should be used}.  They cannot recover a
decisive record that is absent from the accessible set.  Conversely, simply
repeating a poisoned question can retrieve additional replicas of the same
attacker-controlled claim.

We study the intermediate setting in which initial evidence is conflicting,
decisive independent evidence exists in a searchable corpus, and only a small
query budget is available.  Socrates-RAG treats the current answer candidates
as hypotheses, identifies one unresolved necessary premise, and searches for a
record that would discriminate them.  If no such evidence is found, the system
abstains.  This separates evidence \emph{acquisition} from evidence
\emph{aggregation} and connects poisoning defense to selective prediction
\cite{geifman2017selective,xin2021abstention,srinivasan2024selective}.

The strongest alternative is not a passive or single-prompt baseline.  A
capable language model asked for ordinary query expansion may itself infer the
missing premise and authoritative source.  We therefore compare against both
repeated-question retrieval and same-backbone query expansion under exactly
matched search and answer budgets.  This comparison is also a falsification test: if
expansion produces the same actions, the Socratic decomposition has not
established a distinct capability.

Our contributions are:
\begin{itemize}
  \item We formulate active poisoning defense as a finite-budget evidence
        acquisition and selective-disposition problem, introduce a structured
        premise-directed policy, and state a conditional guarantee relative to
        repeated or topical-query policies.
  \item We construct a paired counterfactual active-evidence benchmark with
        hidden evaluator metadata, coordinated replicas, one- and two-hop
        decisive evidence, a matched same-backbone query-expansion control,
        and released-artifact fixed-context references.
  \item On 48 disjoint evaluation worlds, Socrates-RAG improves safe accuracy
        by 14.6 points (8 wins, 1 loss, 39 ties; exact $p=.0391$); trace and
        stratified analyses localize the difference to two-hop premise
        substitution.
\end{itemize}

\section{Related Work}
\label{sec:related}

\paragraph{RAG reliability and poisoning.}
RAG quality depends on retrieval relevance and the generator's use of
context~\cite{gao2023ragsurvey}.  Models can be distracted by irrelevant
or misleading context~\cite{shi2023distracted,yoran2024robustcontext}, and
the relative value of parametric and retrieved knowledge varies by query
\cite{mallen2023trust}.  PoisonedRAG formulates corpus corruption as an
optimization problem~\cite{zou2025poisonedrag}; more recent benchmarking
covers a broader collection of attacks and defenses
\cite{zhang2025benchmarking}.  RobustRAG and TrustRAG represent certified
aggregation and anomaly-filtering approaches, respectively
\cite{xiang2024robustrag,zhou2025trustrag}.  Our protocol is complementary:
it focuses on explicit relations among claims and sources, but the current
study does not establish superiority over these methods under their native
objectives.

ReliabilityRAG makes a different and particularly relevant assumption:
retrieval rank or an explicit reliability weight carries information about
source quality. It constructs a contradiction graph over isolated answers and
selects a rank-aware maximum independent set, with a weighted sampling variant
for larger retrieval sets~\cite{shen2025reliabilityrag}. Its evaluation
deliberately instantiates this assumption through position-controlled attacks
and cleaned answer-bearing corpora. We adopt the same methodological principle:
a mechanism experiment should expose the condition under which the mechanism
can work, while retaining controls where that condition is absent.

\paragraph{Retrieval correction and self-verification.}
Self-RAG learns when to retrieve and critique through reflection tokens
\cite{asai2024selfrag}; Chain-of-Verification decomposes an answer into
verification questions~\cite{dhuliawala2024cove}; CRAG evaluates retrieved
content and invokes corrective retrieval~\cite{yan2024crag}; and FLARE
actively retrieves during generation~\cite{jiang2023flare}.  RARR revises
model outputs using researched evidence~\cite{gao2023rarr}. These systems
motivate explicit verification. Our study isolates a different design choice:
how a model selects the next query when all compared policies share retrieval,
answering, and release components.

Controlled poisoning studies further distinguish adversarial, untouched, and
\emph{guiding} contexts, where guiding passages explicitly supply correct
evidence capable of counteracting an attack~\cite{su2024adversarialrag}. This
distinction motivates our active-rescue regime: the initial context is
insufficient, but decisive evidence exists in the searchable corpus and can be
acquired within a fixed query budget. RAGuard complements this controlled view
with naturally occurring supporting, misleading, and unrelated evidence
\cite{zeng2025raguard}, whereas our benchmark controls evidence reachability
and the query sequence needed to recover it.

\paragraph{Cross-examination and multi-agent debate.}
LM-vs-LM detects factual inconsistencies through cross-examination
\cite{cohen2023lmvslm}.  Multi-agent debate can improve factuality and
reasoning~\cite{du2024debate,liang2024mad}, while theoretical and empirical
analyses show that outcomes depend on participant behavior and debate
protocol~\cite{estornell2024multillm}.  Socrates-RAG differs from open-ended
debate by using bounded roles and machine-readable handoffs.  Our prompt
decomposition uses cross-examination to select retrieval actions rather than
as role rhetoric or an unconstrained discussion protocol.

\paragraph{Factuality, provenance, and selective prediction.}
FActScore evaluates atomic factual precision~\cite{min2023factscore}; ALCE
studies citation generation and citation quality~\cite{gao2023alce}; RAGAS
proposes automated RAG evaluation~\cite{es2023ragas}; and a lightweight
NLI-based provenance checker traces unsupported outputs to context chunks
\cite{sankararaman2024provenance}.  SelfCheckGPT and verbalized uncertainty
provide additional approaches to detecting or expressing uncertainty
\cite{manakul2023selfcheckgpt,lin2022uncertainty,kadavath2022know}.  Our
work adopts the selective-prediction view that reliability must be measured
jointly with answer coverage, not by accuracy alone.

\section{Active-Evidence Socratic Inquiry}
\label{sec:active-method}

\paragraph{Problem.}
For a question $q$, the system first observes a ranked context
$D_0$. It may issue at most $B$ search queries to a fixed corpus
$\mathcal{C}$; query $a_t$ returns the top-$r$ documents
$R(a_t)$. After history
$H_t=(D_0,a_1,R(a_1),\ldots,a_t,R(a_t))$, the system either searches
again, answers, or abstains. The attacker may place query-relevant documents
in $\mathcal{C}$ and replicate one source through multiple pages, but cannot
change the search function, budget, system prompt, or evaluator.

This setting separates two questions that fixed-context evaluation conflates.
A filtering defense asks which members of $D_0$ should be trusted. Active
inquiry additionally asks which unresolved premise should be investigated and
which query is most likely to retrieve evidence that discriminates competing
answers. No policy can recover a fact absent from its accessible corpus; our
method claim is therefore restricted to \emph{rescueable} cases in which
decisive evidence exists but is not present in $D_0$.

\paragraph{Socratic policy.}
The Assertor represents current answer candidates and their necessary
premises. Elenchus identifies one premise whose resolution would separate the
candidates and generates a query containing distinctive entities, docket
numbers, rules, or source names from the evidence. Retrieved pages are added
to the evidence state, after which Elenchus may issue one further query.
Aporia decides whether the acquired evidence supports a unique answer or still
leaves an irreducible conflict. Dialectic returns a short answer with document
citations, or \textsc{CannotDetermine}. A deterministic controller enforces
the budget, de-duplicates documents, and records every terminal error.

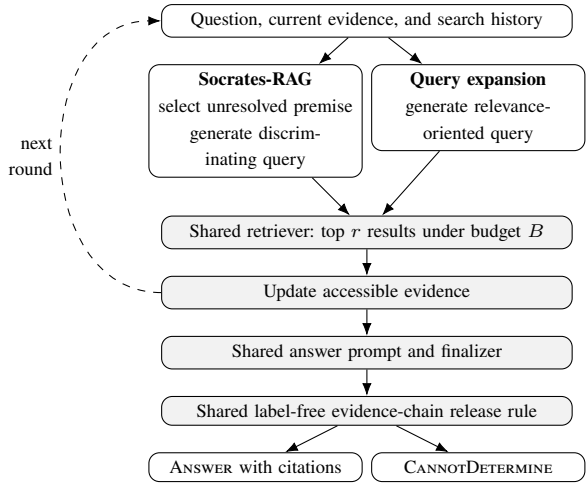
\begin{figure}[t]
\centering
\begin{tikzpicture}[
  node distance=3.5mm,
  >=Latex,
  every node/.style={font=\scriptsize},
  box/.style={
    draw,
    rounded corners,
    align=center,
    inner sep=3pt,
    text width=.62\columnwidth
  },
  query/.style={
    draw,
    rounded corners,
    align=center,
    inner sep=3pt,
    text width=.31\columnwidth
  },
  shared/.style={
    box,
    fill=black!5
  }
]
\node[box] (state) {Question, current evidence, and search history};
\node[query, below=of state, xshift=-.175\columnwidth] (soc)
  {\textbf{Socrates-RAG}\\select unresolved premise\\generate discriminating query};
\node[query, below=of state, xshift=.175\columnwidth] (exp)
  {\textbf{Query expansion}\\generate relevance-oriented query};
\node[shared, below=5mm of soc, xshift=.175\columnwidth] (retrieve)
  {Shared retriever: top $r$ results under budget $B$};
\node[shared, below=of retrieve] (update)
  {Update accessible evidence};
\node[shared, below=of update] (final)
  {Shared answer prompt and finalizer};
\node[shared, below=of final] (guard)
  {Shared label-free evidence-chain release rule};
\node[query, below=of guard, xshift=-.175\columnwidth] (answer)
  {\textsc{Answer} with citations};
\node[query, below=of guard, xshift=.175\columnwidth] (abstain)
  {\textsc{CannotDetermine}};

\draw[->] (state) -- (soc);
\draw[->] (state) -- (exp);
\draw[->] (soc) -- (retrieve);
\draw[->] (exp) -- (retrieve);
\draw[->] (retrieve) -- (update);
\draw[->] (update) -- (final);
\draw[->] (final) -- (guard);
\draw[->] (guard) -- (answer);
\draw[->] (guard) -- (abstain);
\draw[->, dashed] (update.west) to[out=180,in=180,looseness=1.25]
  node[left, align=right, text width=.12\columnwidth]
  {next round} (state.west);
\end{tikzpicture}
\caption{Matched active-retrieval comparison. Socrates-RAG and query expansion
differ only in query generation; retrieval, evidence update, finalization, and
the release rule are shared.}
\label{fig:workflow}
\end{figure}

\paragraph{Matched query-policy comparison.}
At each round, both active policies observe the same question, current
documents, and search history. Socrates-RAG returns an unresolved premise and
a query intended to discriminate competing answers; query expansion returns
an ordinary relevance-oriented query from the same backbone. Both actions are
executed by the same retriever with the same budget and top-$r$ cutoff, and
their accumulated evidence is passed to the same answer prompt, finalizer, and
release rule (Figure~\ref{fig:workflow}). The comparison therefore isolates
query policy rather than additional search capacity or a different answerer.

\paragraph{Evidence-chain release guard.}
Before releasing an answer, the controller requires its cited pages to contain
either (i) a controlling final record that directly binds the entity and
answer, or (ii) a controlling bridge together with its linked terminal record.
Otherwise it converts the output to abstention. This label-free rule reads
document text and cited aliases only; it has no access to gold answers, poison
labels, regime names, hop counts, or evaluator decisive-record IDs.

\paragraph{Objective.}
Let $Y$ be the gold answer, $\bot$ abstention, and $\hat{Y}_{\pi}$ a policy's
output. We report
\begin{align}
\mathrm{SafeCov}(\pi)
  &= \Pr[\hat{Y}_{\pi}=Y],\\
\mathrm{Unsafe}(\pi)
  &= \Pr[\hat{Y}_{\pi}\notin\{Y,\bot\}],
\end{align}
over the fixed population. The mechanism measure is decisive-evidence recall
within budget. A Socratic advantage requires both higher safe coverage and
higher decisive-evidence recall without higher unsafe-answer rate; increased
abstention alone is not a success.

\paragraph{A conditional rescue guarantee.}
Call a state $(q,D_0,\mathcal{C})$ $(B,r)$-separable when (i) a set of
decisive records $E^\star$ determines one answer, (ii) there is a sequence of
at most $B$ queries obtained by resolving necessary premises whose top-$r$
union contains $E^\star$, and (iii) repeating or topically paraphrasing $q$
does not retrieve $E^\star$ within the same budget.  If the final decision
rule returns the evidence-entailed answer whenever $E^\star$ is acquired and
otherwise abstains, a premise-directed policy has safe coverage one and
unsafe-answer rate zero on every separable state, whereas a repeated-question
policy has safe coverage zero.  The proof is immediate from (ii), followed by
the decision rule; (iii) excludes $E^\star$ from the repeated-question
history.  More generally, if a shared finalizer is correct with probability
$\alpha$ after decisive acquisition and both policies abstain otherwise, then
their safe-coverage difference is
$\alpha\,[p_{\mathrm{soc}}(E^\star)-p_{\mathrm{base}}(E^\star)]$.
This is a conditional mechanism guarantee, not a claim that arbitrary Web
states are separable; the experiments measure its premise directly through
decisive-evidence recall. The theorem compares premise-directed inquiry with
repeated or topical-query policies; whether an adaptive same-backbone query
expansion model discovers the same premise is an empirical question.

\paragraph{Replica-robust retrieval lemma.}
Consider a deterministic top-$r$ retriever with score $s(a,d)$.  For a premise
query $a^\star$, let the decisive record $d^\star$ have rank at most $r$, and
assume every replica generated from the attack lineage has score strictly below
$s(a^\star,d^\star)$.  Adding any number of such replicas cannot remove
$d^\star$ from the top $r$: none can increase the number of documents ranked
above it.  In contrast, if a topical query $a_q$ has at least $r$ attack
replicas scoring above $d^\star$, then $d^\star$ is excluded.  Applying the
argument inductively to each revealed bridge proves the analogous statement
for at most $B$ hops.  The substantive policy question is therefore whether
the model generates $a^\star$; our same-backbone expansion baseline directly
tests, and frequently satisfies, that condition.

\section{Counterfactual Active-Evidence Evaluation}
\label{sec:active-benchmark}

We construct counterfactual worlds around fictional people, archives, vaults,
signed curatorial dockets, and storage rules. Fictional values prevent
parametric memory from supplying the answer while allowing exact control over
which evidence is reachable. Each evaluation world begins with a provisional
notice and coordinated pages supporting the same false answer. The decisive
official record exists in the searchable corpus but its identifier is hidden.
In one-hop worlds that record directly resolves the question. In two-hop
worlds it instead reveals an archive or rule that must be inserted into a
second query to retrieve the terminal storage record. Multiple attack pages
share one lineage and therefore do not constitute independent support.

The primary evaluation contains 48 worlds disjoint from method-development
worlds and balanced between 24 one-hop and 24 two-hop cases. Socrates-RAG and
query expansion receive identical initial documents and search histories, use
the same deterministic lexical corpus, and share a two-query, top-three
budget. They also share DeepSeek-V4-Flash at temperature zero, the answer
prompt, and the evidence-chain release rule. The latter was specified on
development worlds before evaluating this set. Released-artifact adaptations
of RobustRAG keyword aggregation and TrustRAG receive the identical initial
documents and backbone route but perform no additional search; neither is
trained, calibrated, or threshold-selected on these worlds. Runtime inputs
expose only the question and retrieved documents; gold answers, attack answers,
hop counts, lineages, decisive-record IDs, and oracle queries remain in
evaluator-only metadata. Provider and parser failures remain in the
denominator.

\section{Results}
\label{sec:active-results}

\paragraph{When expansion is already sufficient.}
Two separate 12-world development analyses delineate the claim. When initial
documents expose exact docket and rule identifiers, Socrates-RAG solves 12/12
active-rescue cases and query expansion solves 11/12: copying the visible
identifier nearly closes the gap. When documents expose only the authoritative
source type, both solve 11/12. Expansion queries such as ``official signed
curatorial registry'' show that a capable backbone can sometimes infer the
missing premise without an explicit Socratic interface. These exploratory
analyses are not pooled with the disjoint evaluation below; they identify the
boundary at which premise-directed structure has little room to help.

\paragraph{Primary comparison.}
Socrates-RAG answers 45/48 evaluation worlds correctly and safely (93.8\%),
compared with 38/48 (79.2\%) for matched query expansion
(Table~\ref{tab:active-rescue}). The paired outcome is 8 wins, 1 loss, and 39
ties; a two-sided exact sign/McNemar test gives $p=.0391$. Decisive-evidence
recall has the same 45/48 versus 38/48 split, while unsafe answers are 0/48
versus 1/48. The correspondence between acquisition and safe correctness
supports the proposed mediator: the gain comes from reaching the evidence
chain rather than answering more aggressively.

\begin{table}[t]
\centering
\small
\setlength{\tabcolsep}{3pt}
\begin{tabular}{lccccc}
\toprule
Method & Safe & Recall & Unsafe & 1-hop & 2-hop\\
\midrule
Socrates-RAG & 45/48 & 45/48 & 0/48 & 24/24 & 21/24\\
Query expansion & 38/48 & 38/48 & 1/48 & 24/24 & 14/24\\
\midrule
RobustRAG keyword & 0/48 & 0/48 & 48/48 & 0/24 & 0/24\\
TrustRAG & 0/48 & 0/48 & 24/48 & 0/24 & 0/24\\
\bottomrule
\end{tabular}
\caption{Results on the same 48 disjoint active-rescue worlds. The top block
contains active acquisition policies; the bottom block contains
released-artifact fixed-context adaptations. Safe denotes a correct released
answer; incorrect released answers are unsafe.}
\label{tab:active-rescue}
\end{table}

\paragraph{Fixed-context boundary.}
On the same 48 worlds, RobustRAG keyword aggregation and TrustRAG acquire no
decisive records and produce no correct answers. RobustRAG releases an
incorrect answer in 48/48 cases; TrustRAG releases 24/48 incorrect answers and
abstains on the remaining 24. This is the expected failure mode when decisive
evidence is absent from the initial set and fictional facts make memorized
knowledge uninformative. These rows directly expose the acquisition boundary
but are descriptive: the paired significance test remains restricted to
Socrates-RAG versus query expansion, and the result is not a claim against
either static defense under its native threat model.

\paragraph{Premise substitution explains the difference.}
Both policies solve all 24 one-hop worlds. The gap is entirely in the 24
two-hop worlds, where Socrates-RAG solves 21 and expansion solves 14. In seven
of eight Socratic wins, the first query retrieves a signed curatorial record
and the second query substitutes its newly resolved archive or rule to find
the terminal storage record. Expansion instead continues with broad
person--archive--vault terms and retrieves coordinated replicas. In the single
paired loss, Socrates-RAG repeats the first premise while expansion broadens to
the storage record. The traces therefore support a specific sequential
mechanism, not a universal advantage for structured prompting.

\paragraph{Cost.}
Both active policies use 144 model calls. Socrates-RAG consumes 111,577 input
and 121,771 output tokens, compared with 101,299 and 66,346 for expansion;
cumulative provider latency is 1,215.6 versus 758.9 seconds. Its accuracy gain
therefore comes at substantial output and latency cost.

\section{Discussion and Limitations}
\label{sec:discussion}

The results support a deliberately narrow claim: premise-directed acquisition
helps when resolving an intermediate premise changes the next useful query.
They do not show an unconditional benefit over a capable expansion model.
Indeed, the development boundary analyses show near parity when identifiers
are visible and parity when the backbone independently infers the appropriate
source type. Socrates-RAG's value is concentrated where newly retrieved
evidence must be composed into a second action.
The same-world static rows sharpen, rather than broaden, this interpretation:
fixed-context aggregation and filtering cannot recover records outside the
accessible set, but remain complementary once those records have been
acquired.

Several limitations bound this conclusion. The evaluation uses fictional,
templated worlds to exclude memorized answers and control evidence
reachability; it does not estimate the prevalence of rescueable conflicts on
the open Web. Retrieval is deterministic and lexical, while poison pages are
coordinated topical replicas rather than attacks adaptively optimized against
the Socratic policy. All active arms use one proprietary model route, so the
result may depend on that model's instruction following. The formal guarantee
is conditional on premise-query separability, which may fail when decisive
evidence is absent, poorly indexed, or requires more than two hops. Finally,
the structured policy substantially increases generated tokens and latency.
Evaluation with public multi-hop corpora, live retrievers, stronger adaptive
attacks, and additional backbones is needed to determine external validity and
the practical quality--cost tradeoff.

\section{Conclusion}

When decisive evidence is missing from the initial context, reliable RAG
depends on choosing what to retrieve next. Socrates-RAG makes that choice by
identifying an unresolved premise, searching for evidence that discriminates
competing answers, and substituting the resolved premise into later queries.
Against a same-backbone expansion control, its benefit appears not on one-hop
retrieval but on two-hop evidence chains that require this substitution. The
result reframes a useful part of poisoning defense as active evidence
acquisition: aggregation protects what is already in context, while
premise-directed inquiry can recover what is absent but reachable. The
same-world RobustRAG and TrustRAG results make this boundary explicit without
turning it into a claim of general superiority over fixed-context defenses.

\clearpage
\bibliography{references}

\clearpage
\appendix
\section{Formal Details}
\label{app:formal}

\paragraph{Active evidence state.}
An instance is $(q,D_0,\mathcal C,R,B,r)$, where $q$ is the question,
$D_0$ the initial context, $\mathcal C$ a fixed searchable corpus, $R(a)$
the deterministic top-$r$ result for query $a$, and $B$ the query budget.
After action sequence $a_{1:t}$, accessible evidence is
\[
D_t=D_0\cup\bigcup_{j=1}^{t}R(a_j),\qquad t\leq B.
\]
A terminal policy returns an answer in $\mathcal Y$ or abstention $\bot$.
A decisive set $E^\star\subseteq\mathcal C$ uniquely entails the gold answer
under the fixed finalizer and release rule.

\paragraph{Premise-query separability.}
An instance is $(B,r)$-separable with respect to baseline class
$\Pi_{\rm top}$ when: (i) some sequence of at most $B$ premise queries
retrieves all of $E^\star$ in its top-$r$ union; (ii) each query after the
first may depend on a premise resolved by earlier retrieved evidence; and
(iii) no policy in $\Pi_{\rm top}$ that only repeats or topically paraphrases
$q$ retrieves all of $E^\star$ under the same budget.

\paragraph{Theorem 1 (conditional rescue).}
Suppose the release rule answers correctly whenever $E^\star\subseteq D_t$
and otherwise abstains. On every $(B,r)$-separable instance, a policy that
executes the separating premise-query sequence has safe coverage one and
unsafe rate zero, whereas every policy in $\Pi_{\rm top}$ has safe coverage
zero.

\paragraph{Proof.}
Condition (i) implies $E^\star\subseteq D_t$ for some $t\leq B$, so the
release rule returns the gold answer. Condition (iii) implies that a topical
policy never acquires the complete decisive set; by assumption its finalizer
therefore abstains. Neither policy emits an incorrect answer. \hfill$\square$

\paragraph{Corollary 1 (imperfect finalizer).}
If a shared finalizer answers correctly with probability $\alpha$ after
decisive acquisition and otherwise abstains, define
$p_\pi=\Pr_\pi(E^\star\subseteq D_B)$. Then
\[
\mathrm{SafeCov}(\pi_s)-\mathrm{SafeCov}(\pi_b)
=\alpha\left(p_{\pi_s}-p_{\pi_b}\right).
\]
Thus decisive-evidence recall is the mechanism mediator measured in our
experiments.

\paragraph{Lemma 1 (replica-robust premise retrieval).}
For query $a^\star$, suppose decisive document $d^\star$ has rank at most
$r$ and every document from the coordinated attack lineage scores strictly
below $d^\star$. Adding arbitrarily many such replicas cannot remove
$d^\star$ from top $r$.

\paragraph{Proof.}
Only documents scoring above $d^\star$ can increase the number of items
preceding it. By assumption no added replica does so, hence its rank remains
at most $r$. Applying the argument after each resolved bridge gives the
bounded-hop result. \hfill$\square$

\section{Method and Prompt Contracts}
\label{app:method}

\paragraph{Controller.}
For each round, Socrates-RAG provides the question, current documents, and
search history to Elenchus. Elenchus returns one atomic unresolved premise,
why it is decisive, a discriminating search query, an expected discriminator,
and a stop decision. The controller rejects an empty query, enforces $B=2$,
retrieves at most $r=3$ pages, and de-duplicates by document ID. A shared
answer call then returns a short answer, cited IDs, and residual uncertainty.

\paragraph{Release guard.}
An answer is released only when cited pages contain either a signed final
record that directly binds the entity to the terminal answer or a signed
bridge together with a signed storage record. Otherwise the controller maps
the output to $\bot$. The guard reads no gold answer, attack label, regime,
hop count, or oracle decisive-document field.

\paragraph{Query-expansion control.}
The strongest active baseline receives the identical question and current
documents and returns an ordinary relevance/recall-oriented search query.
It has the same two-query/top-three budget and shared answer prompt, but it is
not asked to identify or substitute an unresolved premise.

The prompts implement these contracts as structured JSON calls. Each role
returns the fields named above; malformed or missing required fields produce
an error rather than a silently repaired action. The finalizer is identical
across active policies.

\section{Counterfactual Active-Evidence Generator}
\label{app:data}

\paragraph{World schema.}
Each fictional world defines a person, true archive, true vault, false
archive/vault pair, signed curatorial docket, and signed storage rule.
Fictional entities prevent memorized real-world facts from determining the
answer. The 48 evaluation worlds \texttt{AE24--AE71} share no scenario or
world identifier with the development worlds used to specify the method and
release rule.

\paragraph{Initial evidence and coordinated poison.}
Every active-rescue context contains a provisional curatorial notice, two
coordinated direct-answer pages supporting the same false archive/vault pair,
and a verification guide. Exact docket and rule identifiers are hidden.
Searchable pages include three additional coordinated direct-answer replicas,
neutral noise, a signed curatorial record, and---for two-hop worlds---a
separate signed storage record.

\paragraph{One- versus two-hop worlds.}
In one-hop worlds the signed curatorial record directly states the archive and
vault. In two-hop worlds it identifies the true archive and a storage-rule
identifier; a second query must retrieve the signed storage assignment.
The evaluation is balanced at 24 worlds of each type.

\paragraph{Search function.}
Search tokenizes lowercase alphanumeric strings. For query $a$ and document
$d$, the score is
\[
s(a,d)=\sum_{w\in\mathrm{uniq}(a)}
\min\{\mathrm{count}_a(w),\mathrm{count}_d(w)\}.
\]
Documents are sorted by decreasing score and then stable document ID. This
controlled lexical environment makes action differences exactly reproducible;
it is not presented as a natural search-engine model.

\paragraph{Runtime isolation.}
Runtime records expose only scenario ID, question, and initial documents.
Gold answer, attack answer, regime, hop count, decisive IDs, oracle queries,
and lineage annotations live in a separate evaluator file joined only after
inference.

\section{Evaluation Protocol}
\label{app:protocol}

\begin{table}[h]
\centering
\small
\caption{Configuration for the 48-world evaluation.}
\begin{tabular}{ll}
\toprule
Item & Value\\
\midrule
Backbone route & DeepSeek-V4-Flash external\\
Temperature / repeats & 0 / 1\\
Worlds & \texttt{AE24--AE71}\\
Regime & active rescue only\\
Active methods & Socratic, query expansion\\
Static methods & RobustRAG keyword, TrustRAG\\
Active search budget & 2 queries, top 3 each\\
Active terminals / calls & 96 / 288\\
Static terminals / calls & 96 / 384\\
Active / static errors & 0 / 0\\
\bottomrule
\end{tabular}
\end{table}

All errors remain terminal observations; retries do not delete
method--scenario tuples.

\section{Detailed Results}
\label{app:results}

\begin{table}[h]
\centering
\small
\caption{Primary 48-world evaluation.}
\begin{tabular}{lcccc}
\toprule
Method & Safe & Recall & Unsafe & Errors\\
\midrule
Socrates-RAG & 45/48 & 45/48 & 0/48 & 0/48\\
Query expansion & 38/48 & 38/48 & 1/48 & 0/48\\
\midrule
RobustRAG keyword & 0/48 & 0/48 & 48/48 & 0/48\\
TrustRAG & 0/48 & 0/48 & 24/48 & 0/48\\
\bottomrule
\end{tabular}
\end{table}

The paired difference is $7/48=14.6$ points. There are 8 Socratic wins,
1 loss, and 39 ties. With $n_d=9$ discordant pairs, the two-sided exact
sign/McNemar value is
\[
p=2\sum_{k=0}^{1}\binom{9}{k}2^{-9}=0.0390625.
\]
The comparison yields a 14.6-point gain, strictly higher decisive recall,
lower observed unsafe count, and $p<.05$.

\begin{table}[h]
\centering
\small
\caption{Hop-stratified mechanism result.}
\begin{tabular}{lccc}
\toprule
Hop & Worlds & Socrates-RAG & Expansion\\
\midrule
One & 24 & 24/24 & 24/24\\
Two & 24 & 21/24 & 14/24\\
\bottomrule
\end{tabular}
\end{table}

The correctness and decisive-recall counts are identical for both strata.
The entire difference is therefore localized to two-hop evidence acquisition.

\paragraph{Discordant traces.}
Socratic wins occur in \texttt{AE29}, \texttt{AE31}, \texttt{AE39},
\texttt{AE53}, \texttt{AE59}, \texttt{AE65}, \texttt{AE67}, and
\texttt{AE71}; its loss is \texttt{AE37}. In seven wins, Socrates-RAG first
retrieves the signed curatorial record and then inserts the newly resolved
archive/rule into a storage query. Expansion repeats broad person/archive/vault
terms and retrieves coordinated poison replicas. In \texttt{AE37},
Socrates-RAG repeats the curatorial premise, whereas expansion broadens to
``storage register'' and succeeds.

\paragraph{Cost.}
\begin{center}
\scriptsize
\setlength{\tabcolsep}{3pt}
\begin{tabular}{lrrrr}
\toprule
Method & Calls & Input tok. & Output tok. & Latency (s)\\
\midrule
Socrates-RAG & 144 & 111,577 & 121,771 & 1,215.6\\
Query expansion & 144 & 101,299 & 66,346 & 758.9\\
\bottomrule
\end{tabular}
\end{center}
The method establishes no cost advantage.

\section{Training-Free Official Baselines}
\label{app:baselines}

We pin RobustRAG commit \texttt{9bc35b2} and TrustRAG commit
\texttt{11dcea0}. Neither method is trained, fine-tuned, calibrated,
or threshold-selected on our benchmark.

RobustRAG uses the released keyword-aggregation variant. Its secure-decoding
variant is unavailable because the API exposes no token logits; we do not
approximate it under the same name. The repository contains no license file,
which we record as an artifact limitation. TrustRAG retains the released
Sup-SimCSE BERT-base checkpoint, K-means/ngram filtering, thresholds,
internal-knowledge stage, and conflict pipeline. We change
hard-coded CUDA inference to CPU and replace only the generator wrapper with
the shared DeepSeek-V4-Flash JSON API; its repository is MIT licensed.

On the same 48 active-rescue worlds, RobustRAG yields 0/48 safe and 48/48
unsafe answers; TrustRAG yields 0/48 safe and 24/48 unsafe answers, abstaining
on the other 24. Both runs complete 48/48 terminals with zero provider or
parser errors, zero search actions, and zero decisive-record recall. These
results establish the fixed-context acquisition boundary. The primary paired
method comparison remains budget-matched query expansion.

\section{Reproducibility Checklist}
\label{app:repro}

\begin{itemize}
  \item Release deterministic world-generation and lexical-search code.
  \item Release runtime, evaluator labels, and search index as separate,
        versioned files.
  \item Release exact prompts, model route, temperature,
        timeout, retry policy, and provider metadata available from the route.
  \item Release the 96 active-policy terminals with 288 provider-call records
        and the 96 static-policy terminals with 384 provider-call records;
        retain failures rather than deleting them.
  \item Release paired scoring, exact-test, hop-stratification, and discordant
        trace scripts.
  \item Identify RobustRAG and TrustRAG versions and document experimental
        adaptations; never label an engineering proxy as official reproduction.
  \item State that the external route does not expose a public model snapshot,
        and that the controlled lexical benchmark is not a natural-Web
        estimate.
\end{itemize}

\end{document}